\documentclass[twoside]{article}
\usepackage{fleqn,espcrc2}

\usepackage{graphicx}

\newcommand{\AmS}{{\protect\the\textfont2
  A\kern-.1667em\lower.5ex\hbox{M}\kern-.125emS}}
\def\aprle{\buildrel < \over {_{\sim}}} 
\title{Non--standard Neutrino Properties
\thanks{Talk given at NOW2000 Workshop, Conca Specchiulla, Otranto 
(Lecce), Italy, 9--16 September 2000}
}

\author{Maurizio Lusignoli
\address{Dipartimento di Fisica, Universit\`a ``La Sapienza'', 
        and INFN, Sezione di Roma, \\
        Piazzale A. Moro 2, I-00185 Roma, Italy}
       }
       
\begin{document}

\begin{abstract}
A discussion of several exotic models and how well they are able to 
describe the data, with particular emphasis on atmospheric neutrinos.
\vspace{1pc}
\end{abstract}

\maketitle

\section{INTRODUCTION}
The observed suppression in the atmospheric muon neutrino flux and its 
dependence on the neutrino pathlength \cite{Kajita,SK-atmos,oth-atmos} 
can be very well 
described by the hypothesis of massive neutrinos and flavour 
$\nu_{\mu} \leftrightarrow \nu_{\tau}$ oscillations. Also the lack of 
solar electron neutrinos \cite{solar} has been often attributed to 
($\nu_{e} \leftrightarrow \nu_X$) oscillations, with \cite{MSW} or 
without matter effects. A third possible evidence for oscillations, 
coming from the LSND \cite{LSND} results, still needs confirmation by 
other experiments, and I will not consider it in this talk.

It is of great interest to discuss if other theoretical frameworks may 
be able to describe (almost) equally well the present data. Such 
scenarios are generally less popular and therefore have been dubbed 
as ``non-standard'' (although I prefer to call them ``exotic''). 
They generally predict a dependence on neutrino energy $E$ and pathlegth 
$L$ that is different from that of the oscillation formula, so 
that it is mandatory to have a large enough range of $L/E$ values to 
be able to really test the models. In the atmospheric neutrino case, 
this points to the importance of using the much larger energy passing 
upward--muon data \cite{LipLus}; for solar neutrinos there is no 
such possibility, and the discrimination is less easy. 

It may be noticed that the larger energy of the neutrinos inducing 
upward--muon events is also decisive to distinguish $\nu_{\mu}$ 
oscillations into $\nu_{\tau}$ from oscillations into sterile 
neutrinos, as suggested in \cite{LiLu} and actually used by the 
Super--Kamiokande collaboration \cite{tau-neutr,Kajita}. 
Sterile neutrinos cannot be considered sufficiently non--standard to 
be discussed in this talk.

\section{Flavour Oscillations}

\begin{figure}[htb]
\includegraphics[width=16pc]{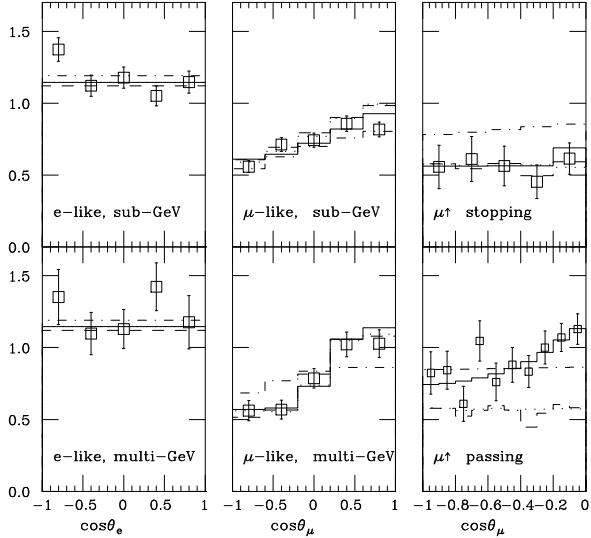}
\vspace{-17pt}
\caption {Ratio data/MonteCarlo for the S--K data 
\protect\cite{SK-atmos}.
The histograms give the best fit predictions for oscillations (solid),
$\nu$ decay (dot-dash), FCNC (dashes) and VEP model (dots).}
\label{plot2.ps}
\end{figure}

Some time ago, we \cite{LipLus} have made a comparison of the 
prediction of several different models with the Super--Kamiokande 
(SK) data \cite{note}. 
A simplified and over--constrained fit was made, in that one common 
normalization parameter was used for six different type of data:
$\mu$--like and $e$--like events, both sub--GeV and multi--GeV, and 
upward--going muons, stopping in the detectors and throughgoing. No 
oscillatory effects were assumed for electron neutrinos, as suggested 
by the CHOOZ \cite{Chooz} results, and the $e$--like events 
were used essentially to fix the normalization. Considering the ratio 
of the observed number of events to the expected (obtained via a 
MonteCarlo calculation), the no-oscillation hypothesis fails to 
describe the data, even assuming independent normalization factors for 
each of the six data sets. In fact, the data show a marked reduction 
(approximately by a factor 2) in the upgoing multi--GeV $\mu$--like 
(semi--)contained events and in the stopping upgoing muons, whose 
energies are similar; the downgoing multi--GeV events are 
unsuppressed; the upgoing passing muons are suppressed by a smaller 
factor, that increases with the pathlength; the sub--GeV muons are 
also suppressed, apparently even if downgoing, although less than the 
upgoing multi--GeV muons. All these results can be well described by 
the $\nu_{\mu}\leftrightarrow \nu_{\tau}$ neutrino oscillations, as 
it may be seen looking at the solid line in fig.\ref{plot2.ps}.

\begin{figure}[htb]
\vspace{9pt}
\includegraphics[width=18pc]{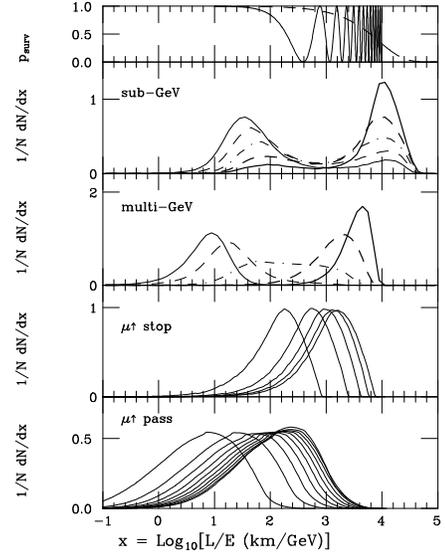}
\vspace{-40pt}
\caption{Distributions in $L/E_\nu$ for the different
event  classes considered.
}
\label{l.ps}
\end{figure}

The reason of the great success of the two--flavour oscillation 
hypothesis in describing the atmospheric neutrino data may be 
understood by looking at Fig.\ref{l.ps}, where the distribution 
in $\log_{10}(L/E)$ for the 25 bins of  
the four sets of $\mu$--like data are plotted in the second to fifth 
panels, and compared to the survival probability given by the 
oscillation hypothesis

\begin{equation}
\label{P-osc}
P^{osc}\,\left({L \over E}\right)  = 1 - \sin^{2}(2\theta)\;\sin^{2}
\left({\Delta m^{2}L \over 4 E}\right) 
\end{equation}

and plotted in the first panel of the figure with the best fit values 
of the parameters: $\Delta m^{2}=3.2\;10^{-3}$ eV$^{2}$ and 
$\sin^{2}(2\theta)=1$.

We note that all the characteristics 
previously mentioned are indeed reproduced. 
The weaker angular correlation between incoming 
neutrino and produced muon in the sub--GeV events explains why even 
the downgoing muons are suppressed: as shown in the second panel of
Fig. 1, a wide 
range of $L/E$ values corresponds to each (muon) angular bin in this 
case, and therefore every bin has some suppression. This is not true 
for the multi--GeV events, where the angular correlation is tighter, 
due to the higher neutrino energy. Finally, the higher energy of the 
passing upgoing muons justifies their smaller suppression and its 
angular dependence, in agreement with the  $L/E$ dependence in 
eq.(\ref{P-osc}).

\section{Exotic Models}

Several exotic models have been proposed in the literature. They are 
not generally able to fit the atmospheric neutrino data, especially 
if the through-- and upward--going muons are included in the data to be 
fitted \cite{LipLus,Fogli}.

A possible effect of a violation of the equivalence principle (VEP) 
\cite{Gasperini} or of a violation of Lorentz invariance 
\cite{Coleman} is a new kind of oscillations among neutrinos, in which 
however the survival probability depends on $L \cdot E$:

\begin{equation}
\label{P-grav}
P^{grav}\,\left({L \cdot E}\right)  = 1 - \sin^{2}(2\theta_{G})\;\sin^{2}
\left({\delta |\phi|\;L\,E}\right) 
\end{equation}

\begin{figure}[htb]
\includegraphics[width=18pc]{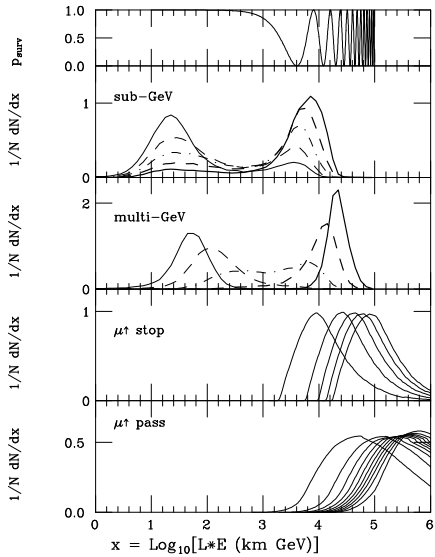}
\vspace{-40pt}
\caption{ Distributions in $L \cdot E_\nu$ for the different
event  classes considered.
}
\label{le.ps}
\end{figure}

In eq.(\ref{P-grav}) $\delta$ is the difference in the gravitational 
coupling of the gravity eigenstates, $\phi$ is the gravitational 
potential and $\theta_{G}$ is the mixing angle that rotates flavor-- 
into gravity--eigenstates. The different dependence on energy and 
pathlength of the survival probability makes a fit to the data much 
worse, even if a fit to the contained events alone is not too bad 
\cite{Foot-VEP}: this may also be visualized considering 
Fig.\ref{le.ps}, the analog of Fig.\ref{l.ps} in 
the present case, that clearly shows the difficulty to obtain for the 
upgoing and passing muons the milder suppression that the data have. 

Another model \cite{Wolf-FCNC} assumes the existence of flavour changing neutral 
currents (FCNC) that may affect the neutrinos crossing the 
earth even if massless.   
If the non--standard elements of the effective potential are  
parameterized as $V_{\mu\tau}=V_{\tau\mu}=\sqrt{2}\,G_{F}\,N_{f} 
\epsilon$ and $V_{\tau\tau}-V_{\mu\mu}=\sqrt{2}\,G_{F}\,N_{f} 
\epsilon^{`}$, with $G_{F}$ the Fermi constant and $N_{f}$ the number 
density of the fermions on which the neutrinos forward--scatter 
($d$--quarks in the calculations discussed), the survival probability 
in this model is given by 

\begin{equation}
\begin{array}{l}
P^{FCNC}\,\left({X_{f}}\right) \; = \\
\ \ \ =\;1 - 
{4\epsilon^{2} \over
{4\epsilon^{2} +\epsilon^{'2}}}\;\sin^{2}
\left[{G_{F}\over {\sqrt 2}}\,X_{f} \sqrt{4\epsilon^{2} + \epsilon^{'2}}
\right] \;,
\end{array}
\label{P-FCNC}
\end{equation}

it is independent on the neutrino energy $E$ 
and it depends only on the column density 
of the fermion $f$ crossed by the neutrino in its path
 $X_{f}=\int^{L}_{0}dx N_{f}(x)$. It may be noted that the 
column density goes to zero abruptly at $\theta$ equal to $\pi/2$, 
contrary to the pathlength which is on the average $L \sim$ 500 km 
for horizontally arriving neutrinos.
 
\begin{figure}[htb]
\includegraphics[width=18pc]{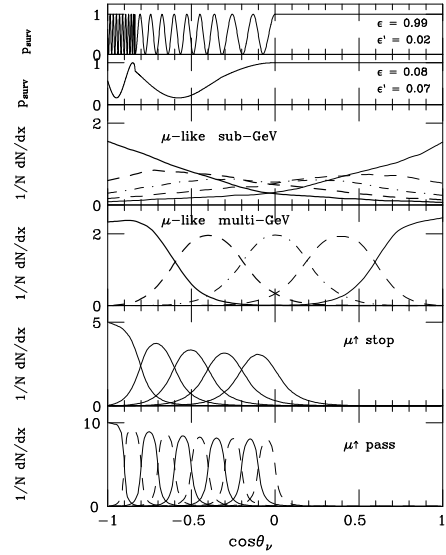}
\vspace{-40pt}
\caption{  Distributions in $\cos \theta_\nu$ for the different
event  classes considered.
In the two top  panels   we show the survival probability
$P(\nu_\mu \to \nu_\mu)$ 
of the two  best  fit  points  as   calculated  in \protect\cite{Valle-FCNC}.}
\label{ct.ps}
\end{figure}

In \cite{Valle-FCNC} two apparently good fits to the contained events 
have been presented. They are however unable to reproduce the upward 
muon data, as it may be illustrated by Fig.\ref{ct.ps}, where the 
distribution in the cosine of zenith angle of the data bins is 
plotted in comparison with the suppression probabilities according to 
the two best solutions of \cite{Valle-FCNC}, plotted in the upper two 
panels.  

Another suggested \cite{Barger} exotic model assumes $\nu_{\mu}$ disappearance 
because of its decay into a lighter neutrino and a majoron. 
The survival probability in this case is 

\begin{equation}
\begin{array}{l}
P^{dec}\,(L/E) = \sin^4 \theta + \cos^4 \theta \; e^{- \alpha 
L/E} +\\
\ \ \ \ + 2
\sin^2\theta \cos^2 \theta \; e^{-\alpha L/2E} \; \cos \left (
\frac{\Delta m^2 L}{2E}\right )
\end{array}
\label{P-decay}
\end{equation}

where $\alpha$ is the ratio of the decaying neutrino mass and 
lifetime. 
The simplest possibility would be pure decay with no mixing, i.e. $\theta=0$,
to which the dashed line in the upper panel of fig.\ref{l.ps} refers. 
It does not reproduce the data satisfactorily. A better solution 
requires mixing, and one has two limiting cases of interest: if 
$\Delta m^2$ refers to the initial and the final mass--eigenstates 
neutrinos in the decay process, then it may be shown that it has to be larger 
than 0.73~eV$^2$, so that the argument of the cosine in 
eq.\ref{P-decay} is very large and it averages to zero. This case has been 
discussed in \cite{Barger} and shown to describe the 
(semi)--contained events reasonably well: it fails however to reproduce
the upgoing muon data \cite{LipLus,Fogli}. 

To summarize the fits made with the models described up 
to this point, I have reported the relevant histograms in fig.\ref{plot2.ps} 
and $\chi^{2}$ values in Table\ref{table:1}.

\begin{table}[htb]
\caption{Statistical significance of the various fits}
\label{table:1}
\newcommand{\m}{\hphantom{1}}
\newcommand{\lra}{\leftrightarrow}
\renewcommand{\arraystretch}{1.2} 
\begin{tabular}{@{}ll}
\hline
Model           & $\chi^{2}$/(d.o.f.) \\
\hline
$\nu_{\mu} \lra \nu_{\tau}$ oscillations   & 33.3/32 \\
VEP           & 143/32 \\
FCNC (with $\epsilon'$=0)    & 149/33  \\
neutrino decay (large $\Delta m^{2}$)  & \m82/32  \\
neutrino decay (no mixing)$^{*}$   & 140/33  \\
\hline
\end{tabular}\\[2pt]
$^{*}$ not plotted in fig.\ref{plot2.ps}.
\end{table}

It may be appropriate to notice that a thorough analysis of the FCNC 
model compared to the most recent data has appeared in the 
meantime \cite{NOFC}, reaching conclusions analogous to ours.

\section{(Still) Successful Exotic Models}
\begin{figure}[htb]
\includegraphics[width=18pc]{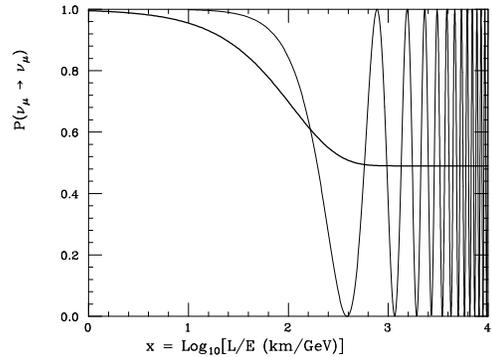}
\vspace{-20pt}
\caption{Survival probabiliity 
for the decay model (heavy solid curve) and $\nu_\mu \leftrightarrow 
\nu_{\tau}$ oscillation model (thin curve).}
\label{fig2.eps}
\end{figure}

The other limiting case among neutrino decay models \cite{Barger+LL}
assumes a very small $\Delta m^2 \aprle 10^{-4}$ eV$^{2}$ 
(indirectly implying the existence of light sterile neutrinos).  
It is a fairly artificial  model, in that it must assume the 
existence of two 
neutrino mass eigenstates with rather large ($\sim 20$ eV) and 
almost equal masses, one of them unstable and the other stable, or 
nearly so. At any rate, it provides a fit to the data which is as 
good as the fit of the flavour oscillation model: 
for $1/\alpha=63\,{\mathrm Km/GeV}$ 
and $\cos^{2}\theta=0.3$ in eq.\ref{P-decay}, one obtains
$\chi^{2}$/(d.o.f.)$= 33.7/32$. The reason of the success should be clear 
looking at fig.\ref{fig2.eps}: the necessary average on $L/E$ implied 
by the broad energy spectrum of atmospheric neutrinos and
by the fact that one only measures energy and angle of the produced muon 
in Super--Kamiokande makes the two distributions hardly distiguishable.

Another successful exotic model \cite{Deco} assumes nonstandard 
Liouville dynamics
\cite{Bena}, leading to $\nu$ decoherence and thus to a damping of
oscillations. The survival probability curve for this model is given 
by the following formula

\begin{equation}
P^{coh} \simeq \textstyle\frac{1}{2}[1+\exp(-\rho L/E)]
\end{equation}

and the best fit is obtained for $\rho\simeq 7\times 10^{-3}$ GeV/Km.
The resulting line, if plotted in fig.\ref{fig2.eps}, is essentially equal 
to the result of the above discussed decay model \cite{Barger+LL}, and 
the fit is therefore equally good. 

A further possibility recently discussed \cite{Barb,Mohap} assumes the 
existence of extra dimensions with (at least one) large radii, and 
sterile, singlet neutrinos propagating in the bulk (like gravitons), while the 
particles that we know are confined to ``the brane". Thus there are 
plenty of sterile states forming Kaluza--Klein towers, to which the muon 
neutrinos could oscillate. The model is however rather constrained, 
and in its minimal version it is not possible to include the LSND result 
\cite{LSND} in the fit. Several possible solutions exist, including 
one in which the muon neutrino oscillate essentially in sterile states 
and there is almost no $\tau$--lepton production. The model parameters 
are rather strongly bound by limits coming from astrophysics 
(supernovae) and cosmology (primordial nucleosynthesis): taking these 
limits at face value, the model is already in trouble \cite{Barb}.

\section{Solar Neutrinos}

As a consequence of the narrow energy range of solar neutrinos, it is 
much more difficult to exclude non--standard solutions of the problem 
in this case. In fact, VEP solutions have been presented, recently 
also for the long wavelength, just--so oscillations \cite{Gago}, and 
they do not seem worse than the flavour oscillation explanation (which 
is admittedly not as good as in the atmospheric case, anyhow).
Also the FCNC model could provide a solution, with somewhat smaller 
(and therefore more acceptable) values of the parameters $\epsilon$ 
and $\epsilon'$ \cite{deHolanda}. The idea of sterile neutrinos moving 
in extra dimensions has been also applied to solar neutrinos
\cite{Dvali} (in fact, before the study of atmospheric neutrinos in 
this kind of model) and shown to provide a possible solution of the 
solar neutrino problem.

An older non--standard explanation of the lack of solar neutrinos 
reaching the earth is based on the existence of transition magnetic 
moments for the neutrinos (of about $10^{-11}$ Bohr magnetons) and is 
due to the so--called resonant spin--flavour precession (RSFP) 
\cite{Lim}. Its predictions depend strongly on the shape of the 
magnetic field distribution inside the sun, largely unknown. The more recent 
analyses \cite{Pulido} show that with a suitable distribution and an average 
magnetic field of $4\div 10$ Tesla a good description of the experimental 
data may be obtained, particularly for Majorana neutrinos. 
The characteristic prediction of this model, namely an anti--correlation 
between the number of sunspots and the flux of solar neutrinos, is not 
supported by the Super--Kamiokande results, however a large magnetic 
field in the interior of the sun could be insensitive to the solar 
activity. Another possible clear signature would be provided by the 
observation of events due to electron antineutrinos, that would be 
produced by the joint effect of RSFP -- inside the sun -- 
changing $\nu_{\mathrm e}$ into, say, $\bar \nu_{\mu}$ and 
normal flavour oscillation -- in the travel to the earth -- inducing 
the  $\bar \nu_{\mu} \to \bar \nu_{\mathrm e}$ transition.

\section{Conclusion}

The data for atmospheric neutrinos, showing the clearest signal of 
physics beyond the (massless neutrino) standard model, are exceedingly 
well described by ``standard" (namely, due to masses and mixing) 
oscillations between $\nu_{\mu}$ and, at least predominantly, $\nu_{\tau}$.
The same data limit severely the exotic options, leaving only some 
very peculiar models as still possible solutions.
Solar neutrinos, on the contrary, have much less power of 
discrimination among models.

Very good resolutions are needed to observe the actual oscillations 
(as opposed to a simple reduction) in the atmospheric case 
\cite{Monolith}. Even for the running (K2K) or future (MINOS,ICARUS)
long--baseline experiments \cite{Hill} the request is demanding. Of course, the 
observation of $\tau$ leptons (by OPERA) could also be of much help.
\vspace{20pt}

I would like to thank the organizers for a very pleasant workshop.



\begin{thebibliography}{99}
\bibitem{note} The fit was made to the early data 
 \protect\cite{SK-atmos}. The use of the most recent data 
 \protect\cite{Kajita} would not change the conclusions.
\end{thebibliography}
\end{document}